\documentclass[prb,10pt,superscriptaddress,twocolumn]{revtex4}
\usepackage{times}
\usepackage{amsfonts} \usepackage{amssymb}
\usepackage{psfrag} \usepackage{graphicx}
\newcommand{\sfrac}[2]{\raisebox{0.095ex}{\scriptsize${\frac{#1}{#2}}$}}
\begin{document}

\title{Gauge-invariant critical exponents for the Ginzburg-Landau model}

\author{H. Kleinert}
\email{kleinert@physik.fu-berlin.de}
\affiliation{ Freie Universit\"at Berlin, Institut f\"ur Theoretische
Physik, Arnimallee 14, D-14195 Berlin }

\author{Adriaan M. J. Schakel}
\email{schakel@boojum.hut.fi}
\affiliation{Low Temperature Laboratory, Helsinki University of
Technology, P.O. Box 2200, FIN-02015 HUT, Finland}

\date{\today}

\begin{abstract}
The critical behavior of the Ginzburg-Landau model is described in a
manifestly gauge-invariant manner.  The gauge-invariant
correlation-function exponent is computed to first order in the $4-d$
and $1/n$-expansion, and found to agree with the ordinary exponent
obtained in the covariant gauge, with the parameter $\alpha=1-d$ in the
gauge-fixing term $(\partial_\mu A_\mu)^2 /2 \alpha$.
\end{abstract}

\maketitle

Despite being one of the most studied field-theoretic models in
theoretical physics, the critical behavior of the Ginzburg-Landau model
is still poorly understood due to nontrivial gauge properties.  The
model is defined by the Hamiltonian
\begin{equation}
\label{H}
\mathcal{H} = \left|(\partial_\mu + i e A_\mu)\phi\right|^2 + m^2
|\phi|^2 + \lambda |\phi|^4 + \frac{1}{4} F^2_{\mu \nu}+ \frac{1}{2
  \alpha} (\partial_\mu A_\mu)^2 ,
\end{equation}
where $F_{\mu \nu} = \partial_\mu A_\nu - \partial_\nu A_\mu$, with
$\mu,\nu = 1, \cdots d$, $e$ and $m$ are electric charge and mass, and
$\lambda$ parametrizes the self-interaction.  The last term with
parameter $\alpha$ fixes the gauge.  The phase transition occurs where
$m^2$ changes sign.  The complex field $\phi$ plays the role of an order
field which has a nonzero expectation value in the ordered phase.

Apart from the standard field-theoretic interpretation, the
Ginzburg-Landau model can be equivalently understood as describing a
random tangle of intertwined electric current loops of arbitrary length
and shape \cite{GFCM,leshouches}.  In the normal state, only a few
current loops are present due to a finite line tension $\theta$.  At the
critical temperature $T_\mathrm{c}$, the tension vanishes and the
current loops become infinitely long.  An important characteristic of
these geometrical objects is their fractal dimension $D$, which at
$T_\mathrm{c}$ is related to Fisher's critical exponent $\eta$,
determining the anomalous dimension of the order field (see below).  In
the absence of gauge fields, this exponent manifests itself in the power
behavior of the correlation function
\begin{equation}
\label{dep}
G(x-x')\equiv \langle \phi(x) \phi^\dagger(x') \rangle
\end{equation}
at the critical point as being $G(x) \sim {1}/{x^{d-2+\eta_\phi}}$.  The
free theory has $G(x) \sim {1}/{x^{d-2}}$, corresponding to $\eta_\phi
=0$.  A nonzero value of the critical exponent $\eta_\phi$ implies that
the dimension of $\phi$ deviates from the canonical, or engineering
dimension $(d-2)/2$.

For a particular value of the Ginzburg-Landau parameter
$\kappa_\mathrm{GL}$, defined by the ratio $\kappa_\mathrm{GL}^2 \equiv
e^2/\lambda$, the Hamiltonian (\ref{H}) is also a dual description of an
ensemble of fluctuating vortex lines of arbitrary length and shape in
superfluid helium \cite{GFCM}, in which case $\phi$ is a
\textit{disorder} field.

In this note, we wish to clarify the properties of this important
exponent, which has been controversially discussed in the past, and very
recently also in the context of quantum electrodynamics (QED)---the
fermionic counterpart of the Ginzburg-Landau model (see
Refs.~[\onlinecite{FTV,Khveshchenko,Ye,GKR}] and references therein).
The poor understanding of this exponent is because in a gauge theory,
the correlation function (\ref{dep}) depends on the gauge parameter
$\alpha$ in Eq.~(\ref{H}).  This has led to a severe theoretical puzzle.
In ordinary local quantum field theories without gauge fields, one can
prove that it must be greater or equal to zero.  In contrast,
renormalization group studies have always produced negative (albeit
gauge-dependent) values, starting with the historic paper by Halperin,
Lubensky, and Ma \cite{HLM}.  In gauge theories, the proof of a
non-negative $\eta$ is not applicable due to the nonlocal nature of the
gauge-invariant correlation function, as has been understood only
recently \cite{Nog}.

We first consider the model close to the upper critical dimension in
$d=4- \epsilon$ dimensions, and extend the Hamiltonian (\ref{H}) for
later discussions to contain $n/2$ complex fields with an
O($n/2$)-symmetric self-interaction.

A one-loop perturbative treatment of the Hamiltionian (\ref{H}) yields
the first term in the $\epsilon$-expansion of the critical exponents.
For the exponent $\eta_\phi$, the well-known result is \cite{HLM,Kang}:
\begin{equation}
\label{eta0}
\eta_\phi = \hat{e}^2_* \,\frac{\alpha - 3}{8 \pi^2} = \frac{6}{n}
(\alpha -3)\epsilon,
\end{equation}
where $\hat{e}^2_* = 48 \pi^2 \epsilon/n$ is the value of the charge at
the fixed point.  For an infrared-stable fixed point to exist in the
two-dimensional space spanned by $e$ and $\lambda$, $n$ must satisfy
$n>12(15 + 4 \sqrt{15}) \approx 365.9$ to first order in the the
$\epsilon$-expansion.  These results are for the massless model ($m=0$).
To avoid infrared divergences, Feynman diagrams are evaluated at finite
external momentum $\kappa$.  Being the only scale available, $\kappa$ is
used to remove the dimension from dimensionful parameters, for example,
$\hat{e}^2 = e^2 \kappa^{d-4}$.  Most calculations reported in the
literature are performed in the Landau gauge ($\alpha=0$) for which
\cite{HLM} $\eta_\phi \to -18 \epsilon/n$.

Since the correlation function (\ref{dep}) is not gauge-invariant, it is
not a physical quantity.  It is therefore not surprising to find that
the critical exponent $\eta_\phi$ depends on the gauge parameter $
\alpha $.  As long as the gauge is fixed by the last term in
Eq.~(\ref{H}), the correlation function is nevertheless well-defined
and, as, following Ref.~[\onlinecite{Lawrie}], can be verified
explicitly to first order in $\epsilon$, the critical exponents satisfy
the usual scaling laws: $\beta_\phi = \frac{1}{2} \nu (d-2 + \eta_\phi),
~ d \nu = \beta_\phi (\delta_\phi + 1)$, with $\eta_\phi$ given in
Eq.~(\ref{eta0}). Since the correlation length exponent $\nu$ is
gauge-independent, $\beta_\phi$ and $\delta_\phi$ depend on $\alpha$.
The exponents $\beta_\phi$ and $\delta_\phi$ specify the vanishing of
the order parameter $\langle \phi\rangle$, respectively for $T
\rightarrow T_\mathrm{c}$ from below and when an external field coupled
linearly to $\phi$ tends to zero.

A physical correlation function must be gauge-invariant, i.e., invariant
under the combined transformations $\phi(x) \to \exp[\mathrm{i} e
\Lambda (x)] \phi(x)$, $A_\mu (x) \to A_\mu (x) - \partial_\mu \Lambda
(x)$.  Such a correlation function is obtained by including a
path-dependent Schwinger phase factor in Eq.~(\ref{dep}), forming
\begin{equation}
\label{indep}
\mathsf{G}(x-x') \equiv \left\langle \phi(x) \phi^\dagger(x')
\mathrm{e}^{-\mathrm{i} e \int_{x}^{x'} \mathrm{d} \bar{x}_\mu
A_\mu(\bar{x})} \right\rangle,
\end{equation}
where the line integral extends from $x$ to $x'$, and the average,
denoted by angle brackets, is taken with respect to the Hamiltonian
(\ref{H}).  At the critical point, we now expect a power behavior
$\mathsf{G}(x) \sim {1}/{x^{d-2+\eta_{\rm GI}}}$, where in contrast to
$\eta_\phi$, the exponent $\eta_{\rm GI}$ has no $\alpha$-dependence.
The exponential in Eq.~(\ref{indep}) can alternatively be written in
terms of an external electric current line $J_\mu(z)$ as
$\mathrm{e}^{-\mathrm{i} \int \mathrm{d}^d z J_\mu(z) A_\mu(z) }$, where
$J_\mu(z) = e \int_x^{x'} \mathrm{d} \bar{x}_\mu \delta(z - \bar{x})$ is
a delta function along the path from $x$ to $x'$ which satisfies the
current conservation law $\partial_\mu J_\mu (z) = e \delta (z- x) - e
\delta (z- x')$, with a current source at $x$ and a sink at $x'$.  In
Refs.~[\onlinecite{KS,KKS}], the gauge-invariant correlation function
(\ref{indep}), with its external current line, was studied in $d=3$ and
found to behave differently in the normal and superconductive state.  In
the normal state, where the line tension $\theta$ of the current line
was shown to be finite, this correlation function decreases
exponentially for large separation, $\mathsf{G}(r) \sim \mathrm{e}^{-
\theta r}$, with $r_\mu = x'_\mu - x_\mu$ being the distance vector.  In
the superconductive state, on the other hand, the line tension vanishes
and the correlation function was found to behave instead as
$\mathsf{G}(r) \sim \exp({e^2 \lambda_\mathrm{L}^2}/{4\pi r})$, with
$\lambda_\mathrm{L}$ being the London penetration depth.  Rather than
tending to zero, the correlation function now reaches a finite value for
large separation.  The finite expectation value at infinite separation
signals that the current lines have lost their line tension and have
become infinitely long.  In the correlation function (\ref{indep}) it
manifests itself in an independence on the path over which the line
integral is taken.  Only the endpoints of the line connecting $x$ and
$x'$ are physical.  The exponent in the correlation function contains a
Coulomb-like interaction between these endpoints.  (Note that the
combination $e \lambda_\mathrm{L}$ is independent of the electric
charge.)

To compute the exponent $\eta_\mathrm{GI}$ to first order in the
$\epsilon$-expansion, the gauge-invariant correlation function
(\ref{indep}) is expanded to order $e^2$.  Then, using Wick's theorem,
three contibutions are obtained besides the lowest order
\begin{equation}
\label{TTT}
\mathsf{G}(r) = G + T_0 + T_1 + T_2,
\end{equation}
containing respectively no, one and two Schwinger phase factors.  The
first contribution $T_0$ is given by the integral
\begin{eqnarray}
\label{T00}
&&\!\!\!\!\!\!\!\! e^2\!\!\! \int\!\! \mathrm{d}^d z \mathrm{d}^d z'
\!\left[\hspace{1pt}\!G(x-z)
\!\!\stackrel{\leftrightarrow}{{\partial}}_{\!z_\mu} \!\! G(z'\!\!-\!x')
\!\!\stackrel{\leftrightarrow}{{\partial}}_{\!z'_\nu} \!\! G(z'\!\!-\!z)
\!\hspace{1pt}\right]\!\! D_{\mu \nu}(z\!-\!z') \nonumber
\end{eqnarray}
where the right minus left derivatives
$\stackrel{\leftrightarrow}{\partial}_{z_\mu}\equiv \partial_{z_\mu} -
\stackrel{\leftarrow}{\partial}_{z_\mu}$ operate only within the square
brackets, and $D_{\mu \nu}$ is the correlation function of the gauge
field $A_\mu$ in (\ref{H}), with the Fourier components
\begin{equation}
\label{aprop}
D_{\mu \nu}(q) = \frac{1}{q^2} \left[ \delta_{\mu \nu} - (1-\alpha)
\frac{q_\mu q_\nu}{q^2} \right] .
\end{equation}
In mometum space this yields
\begin{eqnarray}
\label{T0}
&&\!\!\!\!\!\!\!\!T_0\!=\!e^2\!\!\!  \int \!\!  \frac{\mathrm{d}^d k}{(2
 \pi)^d} \frac{\mathrm{d}^d q}{(2 \pi)^d} \frac{{\rm e}^{\mathrm{i} k
 \cdot r }}{k^4} \frac{ \left(q\! +\! \!  \sfrac{3}{2}k\right)_\mu
 \left(q \!+\!\!  \sfrac{3}{2}k\right)_\nu}{(q\!+\!k/2)^2} D_{\mu
 \nu}(q\!-\!k/2). \nonumber
\end{eqnarray}
 Since the fixed-point value of $e^2$ is of order $\epsilon$ [see below
Eq.~(\ref{eta0})], the integrals can either be evaluated directly in
$d=4$ or using dimensional regularization and taking the limit $d\to 4$
at the end.  Either way gives with logarithmic accuracy
\begin{eqnarray}
T_0(r) = \hat{e}^2 \frac{\alpha-3}{8 \pi^2}\! \int\! \frac{\mathrm{d}^d
k}{(2 \pi)^d} \frac{{\rm e}^{\mathrm{i} k \cdot r }}{k^2}
\ln\frac{k}{\kappa} \!  = - \hat{e}^2 \frac{\alpha\!-\!3}{8 \pi^2} G(r)
\ln \kappa r.  \nonumber
\end{eqnarray}
Adding the free scalar correlation function $G(r)$, we obtain
\begin{equation}
G(r)+ T_0 (r)\! =\! G(r)\! \left[1 \!-\! \hat{e}^2 \frac{\alpha\!-\!3}{8
\pi^2} \ln\kappa r \right]\! \approx\! G(r) r^{- \hat{e}^2
(\alpha\!-\!3)/8 \pi^2}\!,
\end{equation}
which reproduces the old result (\ref{eta0}).

The last term in Eq.~(\ref{TTT}),
\begin{equation}
\label{T2}
T_2(r) = - \frac{1}{2} G(r) \int \mathrm{d}^d z \mathrm{d}^d z' J_\mu
(z) D_{\mu \nu}(z-z') J_\nu (z'),
\end{equation}
factorizes from the start in a scalar and gauge part, with the second
factor---which plays a central role in the study of the gauge-invariant
order parameter of the Ginzburg-Landau model \cite{KS,KKS}---denoting
the Biot-Savart interaction between two segments of the external current
line.  Since the integrals in Eq.~(\ref{T2}) are line integrals, their
value depends on the path chosen.  We choose as integration path the
shortest path connecting the two endpoints, i.e., straight lines and
write
\begin{equation} 
T_2(r) = - \frac{e^2}{2} G(r) \int_0^1 \mathrm{d} u \, \mathrm{d} u'
 r_\mu D_{\mu \nu}[(u'-u)r] r_\nu,
\end{equation} 
after the reparametrization $ \bar{x}_\mu = x_\mu + u r_\mu$,
$\bar{x}'_\mu = x_\mu + u' r_\mu$, with a fixed distance vector $r_\mu =
x'_\mu-x_\nu$ and $0 \leq u, u' \leq 1$.  The integrals are easily
evaluated following Ref.~[\onlinecite{FTV}], with the result
\begin{equation}
- \frac{e^2}{(3\!-\!d) (4\!-\!d)} \frac{\Gamma(d/2\!-\!1)}{4 \pi^{d/2}}
\left[1\!-\! \frac{1}{2} (1\!-\!\alpha) (3\!-\!d) \right] G(r)
r^{4\!-\!d},
\end{equation}
which for $d $ near $4$ yields
\begin{equation}
T_2(r) = \hat{e}^2 \frac{3 -\alpha}{8 \pi^2} G(r) \left[ \ln(\kappa r) +
\frac{1}{\epsilon} \right].
\end{equation}
Due to the appearance of the logarithm multiplying $G(r)$, this gives
the contribution
\begin{equation}
\label{eta2}
\eta_2 = \hat{e}^2_* \frac{\alpha - 3}{8 \pi^2}
\end{equation}
to the Fisher exponent.  When both contributions obtained so far are
subtracted rather than added, one obtains a result (which happens to be
zero) independent of $\alpha$.  As first noted in the the context of QED
\cite{Khveshchenko}, this is because the combination $\eta_\phi -
\eta_2$ characterizes the correlation function
\begin{equation}
\label{check}
{\left\langle \phi(x) \phi^\dagger(x') \right\rangle}{\left\langle \exp
\left(\mathrm{i} e \int_{x}^{x'} \mathrm{d} \bar{x}_\mu A_\mu(\bar{x})
\right) \right\rangle}^{-1},
\end{equation}
which is gauge invariant.

Next, the third, or mixed term in Eq.~(\ref{TTT}), given by the integral
\begin{equation}
\label{T1}
 e \!\int\! \mathrm{d}^d z \, \mathrm{d}^d z' \left[G(x-z)
\!\!\stackrel{\leftrightarrow}{{\partial}}_{\!z_\mu} G(z\!-\!x') \right]
J_\nu(z') D_{\mu \nu}(z\!-\!z'),
\end{equation}
is evalutated.  We expect an $\alpha$-dependent contribution that
precisely cancels the dependence on the gauge parameter found in
Eqs.~(\ref{eta0}) and (\ref{eta2}).  To extract the term of the form
$G(r) \ln(r)$, we use the approximation, cf.\
Ref.~[\onlinecite{Khveshchenko}],
\begin{eqnarray}
T_1(r)& \approx& e G(r) \int \mathrm{d}^d z \, \mathrm{d}^d z'
\left[{\partial_{z_\mu}} G(z\!-\!x') \!-\!  \partial_{z_\mu} G(x \!-\!z)
\right] \nonumber \\&\times &J_\nu(z') D_{\mu \nu}(z\!-\!z'),
\end{eqnarray}
valid with logarithmic accuracy.  Both terms in the square brackets give
the same contribution.  Partially integrating this expression and using
the identity $\partial_\mu \left({x_\mu x_\nu}/{x^4}\right) =
[({3-d})/{2}] \partial_\nu {x^{-2}}$ in $d=4$, we obtain
\begin{eqnarray}
T_1(r) &=& - e^2\frac{\alpha}{2 \pi^2} G(r) \int \mathrm{d}^d z G(z-x')
\frac{1}{(z-x)^2},
\end{eqnarray}
giving $({\hat{e}^2 \alpha }/{4 \pi^2}) G(r) \ln(\kappa r)$ and thus a
contribution to $ \eta $
\begin{eqnarray}
\label{eta1}
\eta_1 = -\hat{e}^2_* \frac{\alpha}{4 \pi^2}.
\end{eqnarray}
As expected, this contribution precisely cancels the $\alpha$-dependence
in Eqs.~(\ref{eta0}) and (\ref{eta2}).  More specifically, we obtain for
the manifesly gauge-invariant correlation function
\begin{equation}
\label{fi}
\eta_\mathrm{GI} = \eta_\phi + \eta_1 + \eta_2 = -\hat{e}^2_* \frac{3}{4
\pi^2} = - \frac{36}{n} \epsilon .
\end{equation}
This value for $\eta_\mathrm{GI}$ is twice that for $\eta_\phi$ obtained
in the Landau gauge $(\alpha=0)$.  Both results coincide, however, when
$\alpha=-3$.

In the current loop description, the critical exponent $\eta_{\rm GI}$
determines the fractal dimension $D$ of the current lines via
\cite{Hoveetal,perco} $D=2-\eta_\mathrm{GI}$.  With $\eta_{\rm GI} < 0$,
the fractal dimension is larger than that of Brownian random walks for
which $D=2$, implying that the current lines are self-seeking rather
than self-avoiding, which makes them more crumpled than Brownian random
walks.  Although higher-order corrections may well change the sign of
$\eta_{\rm GI}$, nothing in the context of the Ginzburg-Landau model
forbids negative values \cite{KS,Nog}, provided $\eta_{\rm GI} > 2 -d$,
or $D<d$.  In the limiting case $D=d$, the current lines would be
completely crumpled and fill out all of space.

Instead of an $\epsilon$-expansion, we may compute the gauge-invariant
critical exponent $\eta_{\rm GI}$ nonperturbatively in the limit of a
large number $n$ of field components.  Then $ \eta $ can be expanded in
powers of $1/n$ for {\em all\/} $2 < d < 4$.

The leading contribution in $1/n$ generated by fluctuations in the gauge
field is obtained by dressing its correlation function with arbitrary
many bubble insertions, and summing the entire set of Feynman diagrams
\cite{AppelHeinz}.  The resulting series is a simple geometrical one,
which leads to the following change in the denominator of the prefactor
in the correlation function (\ref{aprop}):
\begin{equation}
\label{mod}
q^2 \to q^2 + e^2 \frac{n}{2} \frac{c(d)}{(d-1)} \, q^{d-2} ,
\end{equation}
where the second term dominates the first one for small $q$ in $2 < d<
4$.  In Eq.~(\ref{mod}), $c(d)$ stands for the 1-loop integral
\begin{equation}
\label{cd}
c(d) =\! \int \frac{\mathrm{d}^d k}{(2 \pi)^d} \frac{1}{k^2 (k +p)^2}
\biggr|_{p^2=1} \!\!\!= \!\frac{\Gamma(2-d/2) \Gamma^2(d/2-1)}{(4
\pi)^{d/2} \Gamma(d-2)},
\end{equation}
where analytic regularization is used to handle the ultraviolet
divergences.  To leading order in $1/n$ the value of $\eta_\phi$ for $2
< d< 4$ reads \cite{Hikami,VN}
\begin{equation}
\label{largeN}
\eta_\phi = \frac{2}{n} \frac{4-d -(d-1) [4(d-1) -d \, \alpha]}{(4
\pi)^{d/2} c(d) \Gamma(d/2 +1)},
\end{equation}
which depends on the gauge parameter $\alpha$.  For $d=4-\epsilon$, this
result reduces to Eq.~(\ref{eta0}) obtained to first order in the
$\epsilon$-expansion.  The gauge dependence of $\eta_\phi$ is not always
obvious from the results quoted in the literature as often a specific
gauge is chosen from the start, for example, the Landau gauge
\cite{HLM}, where $\eta_\phi \to -40/\pi^2n$ for $d=3$.

The term (\ref{T2}) with the modified gauge-field correlation function
can be evaluated as before.  To extract the dependence on $\ln(r)$ it
proves useful to replace $q^{d-2}$ with $q^{d-2 + \delta}$ in
Eq.~(\ref{mod}), so that the gauge-field correlation function in the
large-$n$ limit becomes
\begin{equation}
\label{apropN}
D_{\mu \nu}(q) = \frac{2}{n e^2} \frac{d-1}{c(d)} \frac{1}{q^{d-2 +
\delta}} \left[ \delta_{\mu \nu} - (1-\alpha) \frac{q_\mu q_\nu}{q^2}
\right] ,
\end{equation}
and to let $\delta \to 0$ at the end.  This leads to
\begin{equation}
T_2(r) = \frac{1}{n} \frac{8(d-1)}{(4 \pi)^{d/2} c(d) \Gamma(d/2-1)}
\left(1 + \frac{1-\alpha}{d-2} \right) G(r) \frac{r^\delta}{\delta},
\end{equation}
with ${r^\delta}/{\delta} \to {1}/{\delta} + \ln(r)$, and
\begin{equation}
\label{GI}
\eta_2 = - \frac{4}{n} \frac{(d-1)(d - 1 - \alpha)}{(4 \pi)^{d/2} c(d)
\Gamma(d/2)} .
\end{equation}
As a check note that, when subtracted from $\eta_\phi$ given in
Eq.~(\ref{largeN}), this yields an $\alpha$-independent result
characterizing the gauge-invariant correlation function (\ref{check}),
\begin{equation}
\eta_\phi - \eta_2 = \frac{1}{n} \frac{16}{(4 \pi)^{d/2} c(d)
\Gamma(d/2-2)}.
\end{equation}
This expression is negative for all $2 < d< 4$.  Specifically,
$\eta_\phi - \eta_2 =-8/\pi^2n$ for $d=3$ and $-4 \epsilon^2/n
+\mathcal{O}(\epsilon^3)$ for $d=4-\epsilon$.

To calculate the mixed term (\ref{T1}), the gauge-field correlation is
needed in coordinate space.  Fourier transforming Eq.~(\ref{apropN}), we
arrive at
\begin{eqnarray}
D_{\mu \nu}(x)& =& \frac{2}{n e^2} \frac{d-1}{c(d)} \frac{4}{(4
\pi)^{d/2} \Gamma(d/2)} \frac{1}{x^{2 - \delta}} \nonumber \\&\times
&\left[\frac{1}{2} (d-3+\alpha) \delta_{\mu \nu} + (1- \alpha)
\frac{x_\mu x_\nu}{x^2} \right].
\end{eqnarray}
Proceeding as before, we obtain
\begin{eqnarray}
T_1(r) &=& - \frac{\alpha}{n} \frac{16(d-1)}{(4 \pi)^{d/2} c(d)
\Gamma(d/2-1)} G(r) \\&\times & \int \mathrm{d}^d z G(z-x')
\frac{1}{(z-x)^{2-\delta}} = - \eta_1 G(r) \ln(r),\nonumber
\end{eqnarray}
with
\begin{eqnarray}
\eta_1 = - \frac{\alpha}{n} \frac{8(d-1)}{(4 \pi)^{d/2} c(d)
\Gamma(d/2)},
\end{eqnarray}
which, being proportional to $\alpha$, should cancel the
$\alpha$-dependence in $\eta_\phi$ and $\eta_2$.  And indeed, as grand
total we find a result
\vspace{-5mm}
\begin{eqnarray}
\label{grand}   \!\!\!
\eta_\mathrm{GI} = \eta_\phi\! + \eta_1\! + \eta_2\! =\! -\frac{4}{n}
\frac{(d^2 \!+\! 2d\! -\!6) \Gamma(d\!-\!2)}{ \Gamma(2 \!-\! d/2)
\Gamma^2(d/2\!-\!1) \Gamma(d/2)}
\end{eqnarray}
which is independent of the gauge parameter $\alpha$.  Remarkably, for
the $d$-dependent gauge choice (of which our $\alpha=-3$ found in the
$\epsilon$-expansion is a special case) the $\eta_\phi$ of
Eq.~(\ref{largeN}) coincides with $\eta_\mathrm{GI}$ to this order in
$1/n$.  With this gauge choice, the trace of the gauge-field correlation
function vanishes, $D_{\mu\mu}(q)=0$.  Since this observation does not
depend on the matter part of the theory, we expect it to hold also in
QED.  And indeed, the values \cite{Khveshchenko,Ye} for the the two
$\eta$ exponents obtained in first order in $1/n$ in $d=3$ and also in
the $\epsilon$-expansion coincide when $\alpha = 1-d$.  Although
higher-order corrections might change this simple relation between the
two exponents, we speculate that in first order the gauge choice $\alpha
= 1-d$ provides a shortcut for obtaining the gauge-invariant result of
other quantities such as the effective potential and mass
renormalization.

The expression (\ref{grand}) is negative for all $2 < d< 4$, with
$\eta_\mathrm{GI} = -72/\pi^2n$ for $d=3$ and $-36 \epsilon/n
+\mathcal{O}(\epsilon^2)$ for $d=4-\epsilon$.  The latter result is in
accord with Eq.~(\ref{fi}) obtained in perturbation theory.  We repeat
that in the context of the Ginzburg-Landau model, negative values are
allowed, provided $\eta_{\rm GI} > 2 -d$, so that the fractal dimension
$D=2 - \eta_{\rm GI}$ of the current lines is smaller than the dimension
of the embedding space.  A negative value merely indicates that the
current lines are self-seeking rather than self-avoiding.

We thank F. Nogueira for many discussions and for pointing out
Ref.~[\onlinecite{FTV}] which instigated us to finish this paper whose
skeleton was written more than three years ago.

One of us (A.M.J.S.) is indebted to M. Krusius for kind hospitality at
the Low Temperature Laboratory in Helsinki and for funding by the
European Union program Improving Human Research Potential (ULTI III).

\begin{thebibliography}{99}
\bibitem{GFCM} H. Kleinert, {\em Gauge Fields in Condensed Matter\/},
Vol.\ I \,\, Superflow and Vortex Lines, World Scientific, Singapore
1989, pp.~1--744.
\bibitem{leshouches} A. M. J. Schakel in: \textit{Topological Defects in
Cosmology and Condensed Matter Physics}, edited by Yu. Bunkov and
H. Godfrin (Kluwer, Dordrecht, 2000), p.~213.
\bibitem{FTV} M. Franz, Z. Te\v{s}anovi\'{c}, and O. Vafek, eprint:
\texttt{cond-mat/0203333} (2002).
\bibitem{Khveshchenko} D. V. Khveshchenko, eprint:
\texttt{cond-mat/0205106} (2002).
\bibitem{Ye} J.-W. Ye, eprint: \texttt{cond-mat/0205417} (2002).
\bibitem{GKR} V. P. Gusynin, D. V. Khveshchenko , and M. Reenders,
eprint: \texttt{cond-mat/0207372} (2002).
\bibitem{HLM} B. I. Halperin, T. C. Lubensky and S. Ma, Phys. Rev.
Lett.  \textbf{32}, 292 (1974).
\bibitem{Nog} F. S. Nogueira, Phys. Rev. B {\bf 62}, 14559 (2000);
H. Kleinert and F. S. Nogueira, eprint \texttt{cond-mat/0104573} (2001).
\bibitem{Kang} J. S. Kang, Phys. Rev. D \textbf{10}, 3455 (1974).
\bibitem{Lawrie} I. D. Lawrie, Nucl. Phys. B [FS] \textbf{200}, 1
  (1982).
\bibitem{KS} M. Kiometzis and A. M. J. Schakel, Int. J. Mod. Phys. B
\textbf{7}, 4271 (1993).
\bibitem{KKS} M. Kiometzis, H. Kleinert, and A. M. J. Schakel,
Fortschr. Phys. \textbf{43}, 697 (1995).
\bibitem{Hoveetal} J. Hove, S. Mo, and A. Sudb\o, Phys. Rev. Lett. 85,
2368 (2000).
\bibitem{perco} A. M. J. Schakel, Phys. Rev. E \textbf{63}, 026115
  (2001).
\bibitem{AppelHeinz} T. Appelquist and U. Heinz, Phys. Rev. D
\textbf{25}, 2620 (1982).
\bibitem{Hikami} S. Hikami, Prog. Theo. Phy. \textbf{62} (1979) 226.
\bibitem{VN} A. N. Vasil'ev and M. Yu. Nalimov, Teor. Mat. Fiz.
\textbf{56} 15 (1983).  \end{thebibliography}
 \end{document}